\documentclass[aps,prl,reprint,showkeys,showpacs]{revtex4-1}                            

\usepackage{amsmath,amssymb,amsfonts,amsbsy,bm}   
\PassOptionsToPackage{caption=false}{subfig}      
\usepackage{graphicx,graphics,color,epsfig,subfig}
\usepackage[sort&compress]{natbib}                
\usepackage{color}                                

\usepackage{float}                                
\DeclareMathAlphabet{\mathscr}{OT1}{pzc}           
                                 {m}{it}

\DeclareGraphicsExtensions{.eps,.tif}    

\newcommand{\bra}{\langle}
\newcommand{\ket}{\rangle}
\newcommand{\be}{\begin{equation}}
\newcommand{\ee}{\end{equation}}
\newcommand{\ba}{\begin{eqnarray}}
\newcommand{\ea}{\end{eqnarray}}
\newcommand{\ccc}{\lambda}

\begin{document}
\title{Quantum Rabi model for $N$-state atoms}
\author{Victor V. Albert}
\email[]{victor.albert@yale.edu}
\affiliation{Department of Physics, Yale University, P.O. Box 208120, New Haven, CT 06520-8120, USA}
\date{\today}
\begin{abstract}
A tractable $N$-state Rabi Hamiltonian is introduced by extending the parity symmetry of the two-state model. The single-mode case provides a few-parameter description of a novel class of periodic systems, predicting that the ground state of certain four-state atom-cavity systems will undergo parity change at strong coupling. A group-theoretical treatment provides physical insight into dynamics and a modified rotating wave approximation obtains accurate analytical energies. The dissipative case can be applied to study excitation energy transfer in molecular rings or chains.

\end{abstract}
\maketitle

Interactions between spin systems and harmonic oscillators (boson modes) have been studied for over 70 years \cite{rabi, *rabi2, fg_original, *wagner_fg, jc, *Paul1963, *Shore1993}. One of the most well-known, the quantum Rabi model \cite{rabi, *rabi2}, is a phenomenological Hamiltonian describing interactions between a two-level system and a cavity mode. The model has also formed a basis of understanding for exciton-phonon interactions \cite{spin_boson_shore_sander_FG, *wagner} and, along with its multi-mode extension, has numerous established applications in chemistry and physics (see \cite{Viola1998, *dimer_one_moder_small_j_irish, *dimer_one_mode_irish_grwa_prl, braak, me} and refs. therein). The Jaynes-Cummings (J-C) \cite{jc, *Paul1963, *Shore1993} model is obtained by taking the Rabi model in the rotating wave approximation (RWA), where the ``counter-rotating'' terms are ignored (see e.g. \cite{scully}). While the J-C model is sufficient to study small atom-field coupling, the RWA breaks down at large coupling \cite{spin_boson_deep_strong_coupling, Larson2011} and the full Rabi model is needed. Experimental techniques have accessed these strong-coupling regimes \cite{You2011} and there is much ongoing interest in future experimental realizations in both cavity \cite{Schuster2010, *Crespi2011, *Chang2012} and circuit \cite{Longhi2011, *Romero2012} QED.

Many-site spin-boson interaction, e.g. multi-state atom-cavity interaction \cite{Scully2006, *Bianchetti2010, *Srinivasan2011, *squitrid} or excitation energy transfer in multi-chromophoric systems \cite{Kolli2011}, continues to be a subject of significant interest, dictating a need for extensions of the two-state model. Extensions of the J-C model have been studied extensively \cite{Yoo1985, genjc3, Hagelstein2008, quadrumer_general}, but are no longer applicable in the strong-coupling regime. Exciton-phonon generalizations which extend the parity/reflection symmetry of the Rabi system \cite{wagner_fg3, *wagner_trimer, *wagner_gfg} are neither tractable nor applicable to atom-cavity systems. Most importantly, the Rabi model is the single-mode version of a dissipative (infinite-mode) spin-boson model \cite{leggett}, signifying that light-matter interaction is a simplified manifestation of a more fundamental interaction between a two-state system and a dissipative environment. Previous dissipative \cite{Gilmore2005, *Gilmore2008, cheng2009, Egger1994, Jing2010} generalizations have neither extended the symmetry nor preserved this correspondence. Motivated by these properties, this Letter presents a symmetry-preserving $N$-state extension of the Rabi model. The extension includes counter-rotating terms in a rigorous, intuitive, and mathematically manageable way, using a minimal number of parameters and paving the way for applications to multi-level atom-cavity experiments at both weak and strong coupling. A group-theoretical approach \cite{fg_original, *wagner_fg} provides numerical advantages and physical insight into dynamics of the single-mode case. The symmetric generalized RWA \cite{me} is applied to obtain accurate analytical energies/eigenstates valid for strong coupling. The above procedures are significantly simplified via the generalized spin matrices \cite{gsm}, providing a new tool for the treatment of general periodic systems. The corresponding infinite-mode extension can in turn be applied to periodic dissipative $N$-state systems.

\begin{figure}[htpp]
 \includegraphics[width=0.48\textwidth]{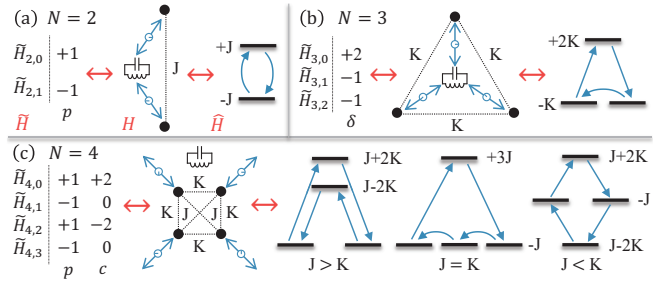}  
 \caption{
    \small (color online) (a) The partially diagonalized (left panel), position (center), and energy (right) representations form three equivalent interpretations of the two-state Rabi model. (b),(c) analogously depict the $N$-state Rabi model from Eq. (\ref{hd}) with $N=3,4$, respectively. The parameters $J,K$ form the energies of the system, $\{p,\delta,c\}$ are each system's conserved quantum numbers, and $\widetilde{H}_{N,n}$ are boson chains (defined in text).
  \label{f1}}
\end{figure}
\normalsize

This work will present the $N$-state Rabi model's physical motivation in two different representations, discussing the $N=3,4$ cases from the viewpoint of atom-cavity physics. A discussion of the conserved quantum numbers, a third symmetry-based representation, and dynamical properties of the single-mode case follows. The remaining space is devoted to a brief description/application of the dissipative case. For reference, the three representations discussed here are graphically outlined in Fig. \ref{f1} for $N=2,3,4$.

\textit{Position representation.---}Consider the Hamiltonian $H=H_{sys}+H_{field}+H_{int}$ describing an $N$-state system interacting with electric and magnetic fields (${\bf E}$ and ${\bf B}$) of a cavity mode of frequency $\omega$ and wavenumber $\mathscr{k}_{\omega}$. The Hamiltonian will be formulated in the {\it position} representation, where the interaction $H_{int}$ is on the (spin) diagonal. This differs from the traditional introduction of the Rabi model in the {\it energy} representation, where the interaction is off-diagonal (pg. 194 of \cite{scully}). While the two formulations are equivalent for the two-state case, this version provides a natural symmetry-based extension. Assume a class of systems defined by
\be\label{sys}
H_{sys}=\sum_{k}\sum_{n=0}^{N-1}J_{k}\left(|n\ket\bra n+k|+|n+k\ket\bra n|\right),
\ee
where $\left\{|n\ket\right\}_{n=0}^{N-1}$ form a complete set of position eigenstates and $k$ sums over all neighboring sites. Transforming $H$ into the energy representation, i.e., diagonalizing $H_{sys}$, would obtain an $N$-state ``atom'' with energies determined by the parameters $J_{k}$. The interaction is
\be\label{coup}
H_{int}=-\mathbf{d}_{E}\cdot\mathbf{E}-\mathbf{d}_{B}\cdot\mathbf{B},
\ee
where $\mathbf{d}_{E(B)}$ is the electric (magnetic) dipole moment operator. Assuming the fields are constant in time over the neighborhood of the atom, the cavity mode can be quantized \cite{scully} with $\mathbf{E}\propto(a+a^{\dagger})\sin(\mathscr{k}_{\omega}z)$, $\mathbf{B}\propto i(a^{\dagger}-a)\cos(\mathscr{k}_{\omega}z)$, and $H_{field}=\omega a^{\dagger}a$ (with $a$ and $a^\dagger$ denoting creation and annihilation operators of the mode). Switching $\sin \leftrightarrow \cos$ by introducing $b=ae^{-i\frac{\pi}{2}}$, discretizing the $z$-axis over the $N$ position states of the atom ($\mathscr{k}_{\omega}z|n\ket=\frac{2\pi n}{N}|n\ket$), and relegating the coupling strengths to a parameter $\ccc$ obtains the $N$-state Rabi Hamiltonian \footnote{The main assumption of the model could in principle be manifested by aligning the respective dipole moments with the fields in such a way that the couplings contribute equally. Additionally, purely electric and magnetic dipole forms of $H_{int}$ are equivalent for the two-level case, as is the general rule [A. E. Siegman, {\it Lasers} (University Science Books, Mill Valley, California, 1986) p. 1222]. This approach predicts that this is not the case for higher-level systems.}
\ba \label{pr_hse2}
H &=& \omega b^{\dagger}b+\ccc\sum_{n=0}^{N-1}(be^{i\frac{2\pi}{N}n}+b^{\dagger}e^{-i\frac{2\pi}{N}n})|n\ket\bra n| \nonumber\\
  & & \,\,\,\,\,\,\,+\sum_{k}\sum_{n=0}^{N-1}J_{k}\left(|n\ket\bra n+k|+|n+k\ket\bra n|\right).
\ea
For the two-state case ($N=2$), this simplifies to the original Rabi model
\be\label{h2}
H_2=\omega b^{\dagger}b+\ccc\left(b+b^{\dagger}\right)\sigma_{z}+J\sigma_{x},
\ee
where $J\equiv J_{\frac{N}{2}}$ and $\sigma_{x,z}$ are the usual Pauli matrices. One can also interpret $H$ as a normal mode smeared over a tunneling $N$-site system (discussed later). As a result, this extension maintains the correspondence between atom-field interaction and a more general spin-boson model.

It will now be shown that re-expressing the model in terms of the generalized spin matrices, the unitary generalization of the Pauli matrices \cite{gsm}, will reduce mathematical complexity while increasing physical understanding. Suppressing dependence on $N$, generalized spin matrices for $0\leq j,k < N$ are defined (via modulo $N$) as
\begin{equation}\label{spin}
S_{j,k}=\sum_{n=0}^{N-1}e^{i\frac{2\pi}{N}nj}|n\ket\bra n+k|=\left(S_{1,0}\right)^{j}\left(S_{0,1}\right)^{k}.
\end{equation}
With the details relegated to \footnote{See Supplemental Material at [URL] for details on the generalized spin matrices, the Fulton-Gouterman \cite{fg_original, *wagner_fg} transformation that partially diagonalizes $H$, and the S-GRWA.}, the reader need only keep in mind the function of the two indices: $j$ determines the phase at each entry $n$ while $k$ determines the entry's location. The matrices $S_{1,0}$ and $S_{1,0}^\dagger$ elegantly express $H_{int}$ while $S_{0,k}+S^\dagger_{0,k}$ describes the neighbor couplings of $H_{sys}$. For $0<k\leq\kappa\equiv\textstyle\left\lfloor \frac{1}{2}\left(N-1\right)\right\rfloor$ (with $\left\lfloor N \right\rfloor$ the floor function), Eq. (\ref{pr_hse2}) is thus re-expressed as
\begin{eqnarray}\label{hd}
\nonumber H&=&\omega b^{\dagger}b+\ccc(b\, S_{1,0}^{\,}+b^{\dagger}S_{1,0}^{\dagger})\\&&\,\,\,\,\,\,\,\,\,\,\,\,\,\,\,\,\,\,\,\,\,\,\,\,+JS_{0,\frac{N}{2}}+\sum_{k=1}^{\kappa}J_{k}(S_{0,k}^{\,}+S_{0,k}^{\dagger}).
\end{eqnarray}

\textit{Energy representation.---}One can now transform $H$ into the energy representation $\widehat{H}=V^\dagger H V$ using the unitary transformation
\begin{equation}\label{v}
V=\frac{1}{\sqrt{N}}\sum_{k=0}^{N-1}e^{i\frac{2\pi}{N}k^{2}}S_{k,k}S_{1,0},
\end{equation}
linking the above formulation with the well-established picture of dipole transitions in $N$-state systems \cite{scully}. The transformed Hamiltonian
\begin{equation}\label{hdr}
\widehat{H}=\omega b^{\dagger}b+\ccc(bS_{2,1}e^{i\frac{4\pi}{N}}+b^{\dagger}S_{2,1}^{\dagger}e^{-i\frac{4\pi}{N}})+\widehat{H}_{sys}
\end{equation}
models an $N$-state atom coupled to a field mode (with the option for more modes \footnote{For the three-level two-mode case, another mode $b_2$ could be added to $H_3$ via $b_2 S_{2,0} + b_2^\dagger S^\dagger_{2,0}$. One can compare to a cascade ($\Xi$) three-level configuration, recently expressed in the position representation in X. Ren, H. Cong, X. Wang, and J. Xia, Sci. China Phys. Mech. Astron. {\bf 54}, 1625 (2011). For arbitrary $N$, a mode $b_n$ can couple via $S_{n,0}$ for $n<N$, providing a non-RWA model of interaction between an $N$-state atom and $N-1$ modes \cite{genjc3} which preserves cyclic symmetry.}). The symmetry of the coupling determines which states are coupled by the mode and the state energies are determined by $\widehat{H}_{sys}$. For $N=2$, Eq. (\ref{hdr}) reduces to $\widehat{H}_2$, the Rabi Hamiltonian in the energy representation [$\sigma_x \leftrightarrow \sigma_z$ in Eq. (\ref{h2})]. The three- and four-state cases are reviewed below.

$N=3$: Setting $J_1\equiv K$ in Eq. (\ref{hd}), the three-state case in the energy representation is
\begin{equation}\label{vh3}
\widehat{H}_{3}=\left(\begin{array}{ccc}
\omega b^{\dagger}b+2K & \ccc be^{-i\frac{2\pi}{3}} & \ccc b^{\dagger}\\
\ccc b^{\dagger}e^{i\frac{2\pi}{3}} & \omega b^{\dagger}b-K & \ccc be^{i\frac{2\pi}{3}}\\
\ccc b & \ccc b^{\dagger}e^{-i\frac{2\pi}{3}} & \omega b^{\dagger}b-K
\end{array}\right).
\end{equation}
The above is a three-level atom with an initially degenerate ground state and energy separation $3K$ coupled to one cavity mode \footnotemark[3]. The states are thus arranged in a $\Lambda$ configuration (with inversion of $K$ obtaining a V configuration), similar to well-studied $\Lambda$-systems \cite{Yoo1985}. However, dipole transitions occur between all three levels while extending the symmetry and maintaining the relative simplicity of the original Rabi model. The bottom left entry in Eq. (\ref{vh3}) describes the process in which the atom makes a transition from the upper to the lower level and a photon is annihilated \cite{scully}. The RWA (with respect to $\omega b^\dagger b+\widehat{H}_{sys}$) removes this transition, relating Eq. (\ref{vh3}) to well-established extensions of the J-C model \cite{Yoo1985}. The coupling between the ground states represents an ac-Stark shift (similar to the J-C model in the dispersive regime \cite{Boissonneault2008, *Zueco2009}, relevant to non-demolition measurements), which interestingly remains relevant after the RWA.

\begin{figure}[htpp]
 \begin{center}
 \subfloat(a){\includegraphics[width=0.4\textwidth]{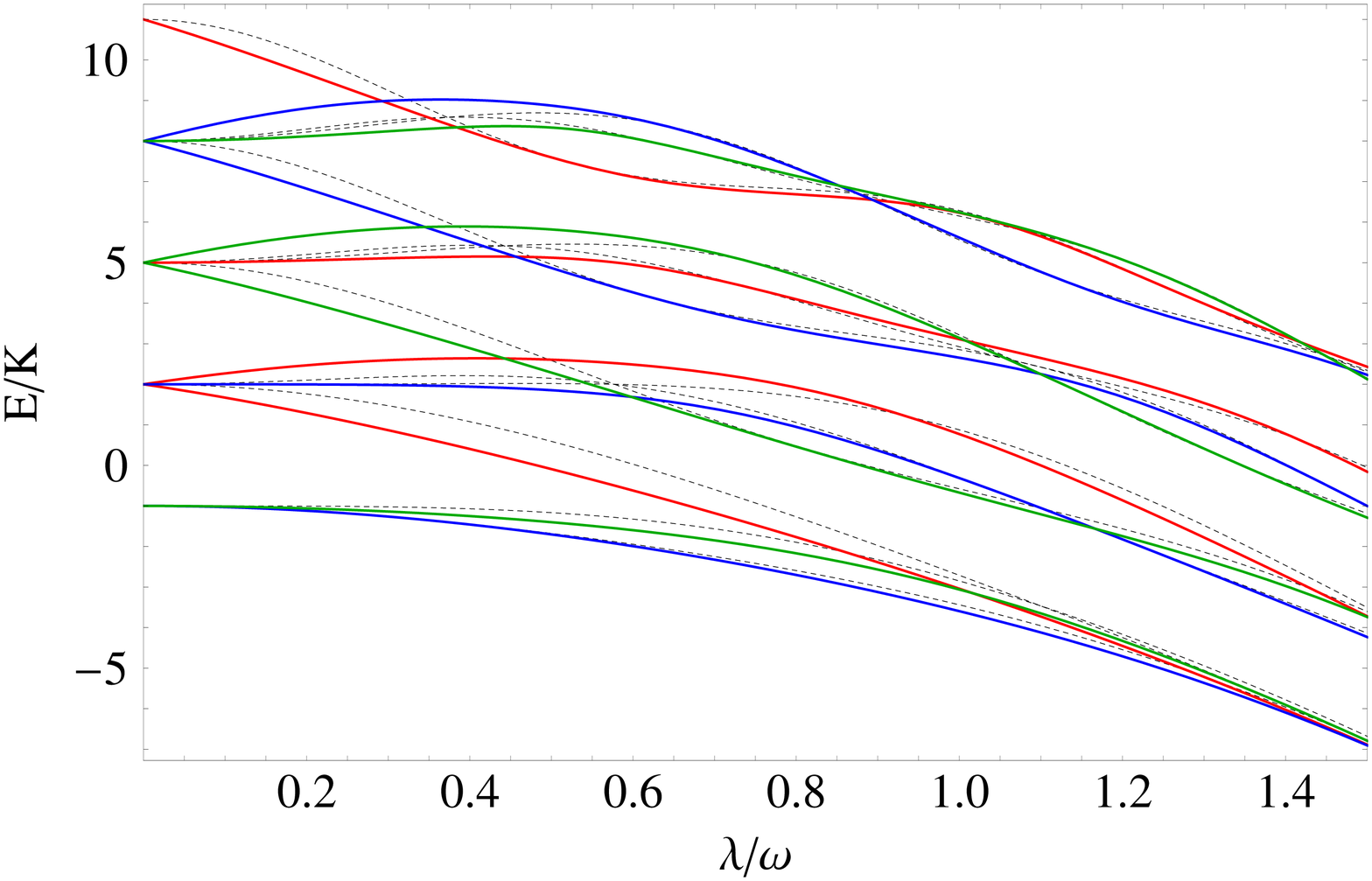}}\\ 
 \subfloat(b){\includegraphics[width=0.4\textwidth]{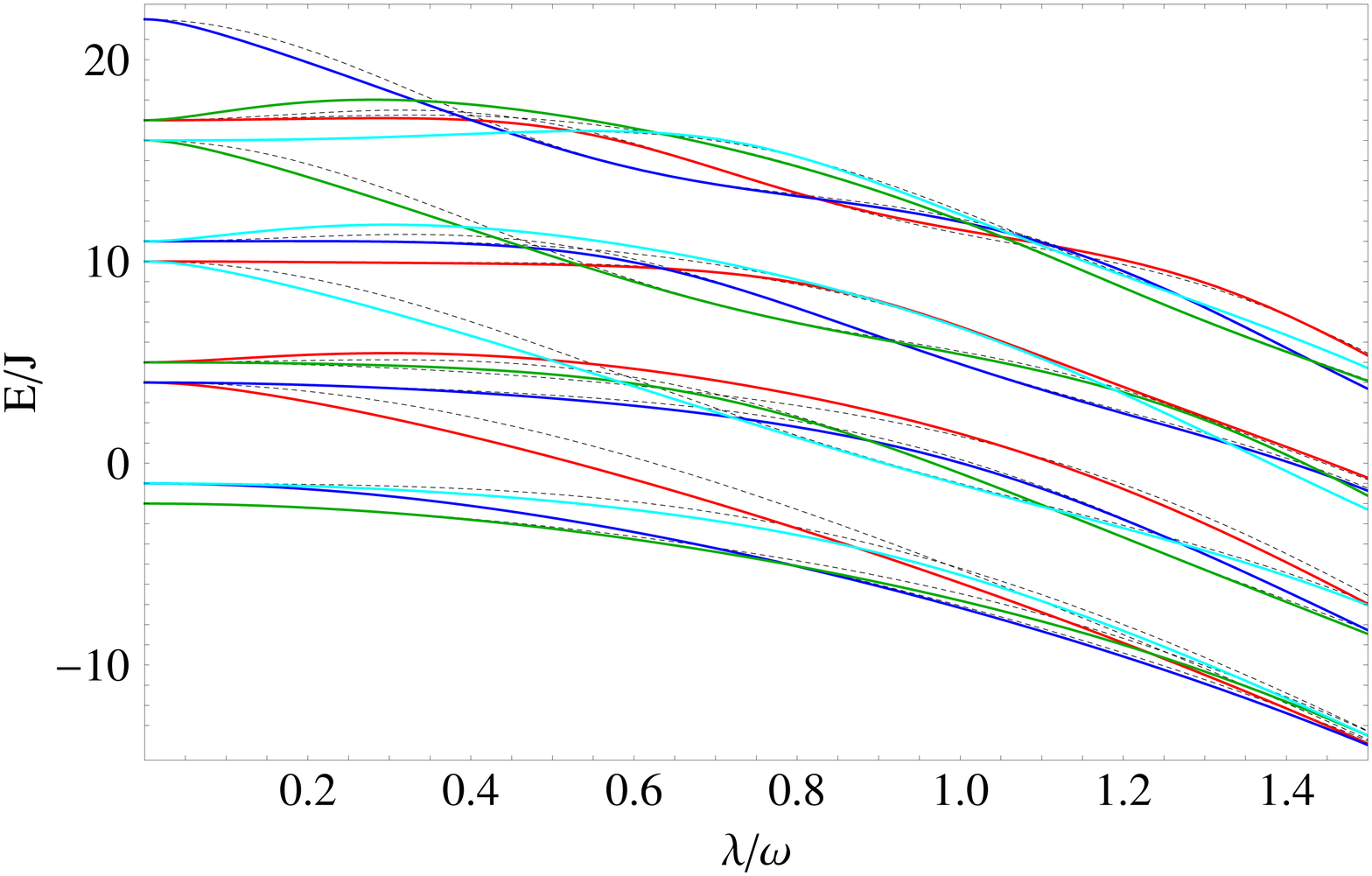}}
 \end{center}
 \caption{
    \small (color online) Correlation diagrams of energy vs. coupling $\ccc$ for (a) $H_3$, a resonant $\Lambda$-configuration with $K=\omega/3$, and (b) $H_4$, a $\Diamond$-configuration with $K=\omega/4$ and $J=\omega/6$ ($\omega =1$). The numerical energies belong to chains $\widetilde{H}_{N,n}$ (for $n < N=3,4$) represented as red, blue, green, and cyan, in that order, while approximate S-GRWA energies are dashed. The model predicts that the $\Diamond$-configuration may have a different ground state at $\ccc/\omega \approx 1$ than at weak coupling. \label{f2}}
\end{figure}

$N=4$: For the four-state case, $\widehat{H}_{sys} = \text{diag}\left\{ J+2K,-J,J-2K,-J\right\}$. Depending on the relation between $J>0$ and $K$ [Fig. \ref{f1}(c)], one can obtain either a double-$\Lambda$, tripod, or $\Diamond$ four-state configuration \cite{quadrumer_general, Fleischhauer2005}. Inversion of $J$ obtains inverted tripod and double-$\Lambda$ configurations; inversion of $K$ leaves the system invariant just like inversion of $J$ for $N=2$.
The cavity frequency $\omega$ can be tuned to the three possible transition frequencies of the atom, producing a four-parameter model for treating single- and (in the $\Diamond$ case) multi-level transitions in several related systems. Additionally, $H_4$ can be separated into two effective two-state systems as $K\rightarrow 0$. One striking feature is that the ground state can change for increasing values of $\ccc$, a property not seen at $N<4$. Shown in Fig. \ref{f2}(b) for a particular $\Diamond$-configuration, the ground state at small coupling (blue) is surpassed by the unperturbed first excited state (green) as the coupling increases.

\textit{Conserved quantum numbers.---}The generalized spin matrices allow one to easily construct the complete set of conserved quantum numbers for $H$, providing important insight into dynamics \cite{spin_boson_deep_strong_coupling} and integrability \cite{braak}. It can be shown that the Hamiltonian (\ref{hd}) possesses an $N$-fold rotational symmetry and commutes with the rotations $\left\{ \mathcal{R}_{n}S_{0,n}\right\} _{n=0}^{N-1}$, where the bosonic rotation $\mathcal{R}_{n}=\exp\left(i{\textstyle \frac{2\pi}{N}nb^{\dagger}b}\right)$ and the parity/reflection $\mathcal{R}\equiv \mathcal{R}_{\frac{N}{2}}$ is present for even $N$. These can be compiled into the general $N$-state commuting operator
\begin{equation}
\mathbf{N}=J\mathcal{R}S_{0,\frac{N}{2}}+\sum_{k=1}^{\kappa}J_{k}(\mathcal{R}_{k}S_{0,k}+\mathcal{R}_{k}^{\dagger}S_{0,k}^{\dagger}),
\end{equation}
consisting of the family of $\kappa$ commuting Hermitian operators multiplied by site couplings $J_k$ (with the additional parity operator for even $N$). For the original $N=2$ case, $\mathbf{N}$ reduces to $J$ multiplied by the well-known spin-boson parity $\sigma_x \mathcal{R}$ \cite{braak, spin_boson_deep_strong_coupling}. This result shows that these $N$-state atom-cavity systems not only preserve parity for any even $N$, but are classified by other quantum numbers for $N>2$. For example, the three-state case contains a conserved quantum number $\delta=2,-1$ while the four-state state system has two: parity $p=\pm1$ and ``cascade'' number $c=0,\pm2$.

\textit{Analytical insight.---}The rotational symmetry of $H$ allows decomposition into $N$ infinite-dimensional subspaces ({\it boson chains}, denoted as $\widetilde{H}_{N,n}$) via a group-theoretic transformation $U$ \footnotemark[2]. In this partially diagonalized representation, the Hamiltonian $\widetilde{H}=U^{\dagger}HU$ is diagonal in the spin subspace with $\bra n^{\prime}|\widetilde{H}|n\ket=\delta_{n^{\prime},n}\widetilde{H}_{N,n}$. These chains are isomorphic to $H$ and provide significant numerical advantages \cite{me}. For the two-site case, $\widetilde{\mathbf{N}} \rightarrow J\sigma_z$, resulting in parity chains \cite{spin_boson_deep_strong_coupling}, shown in the left panel of Fig. \ref{f1}(a). The chains and their respective quantum numbers for the three- and four-state cases are depicted in Fig. \ref{f1}(b) and (c), respectively. The numerical energies for $H_3$ and $H_4$ are plotted in Fig. \ref{f2}(a) and (b), respectively, and each chain is labeled by a color. The spectrum of $H$ demonstrates the familiar braid-like crossing pattern of the two-level Rabi Hamiltonian with the addition of more braids.

The group theoretical approach is also useful for extending analytical approximations, such as the symmetric generalized RWA (S-GRWA \cite{me}, applied in \footnotemark[2]). The S-GRWA energies (dashed in Fig. \ref{f2}) are most accurate in the deep-strong coupling regime ($\lambda \approx \omega$), where the symmetry and chain structure are important \cite{spin_boson_deep_strong_coupling}. While the S-GRWA also fares well at $\lambda \ll \omega$, the symmetry is not terribly relevant in that region and the original RWA may be applied without loss of accuracy. In the weak coupling regime, it is anticipated that classical J-C collapses and revivals \cite{scully} will occur in the system dynamics, but this time between multiple atomic states. The notably different behavior of $H$ in the strong-coupling regime will likely be an extension of that described in \cite{spin_boson_deep_strong_coupling} and will be chain-dependent. Both of these regimes (as well as transitions between them) reveal opportunities for interesting manifestations of both well-known and newly-discovered phenomena of the original two-level case.

\begin{figure}[htpp]
 \includegraphics[width=0.48\textwidth]{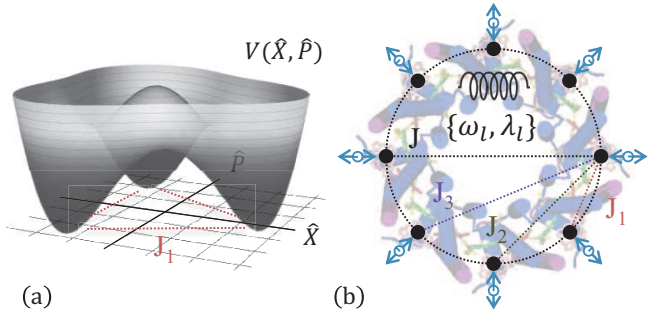}  
 \caption{
    \small (color online) \label{f3}
    In a manner analogous to \cite{leggett}, (a) is a pictorial representation of a symmetric three-well potential in the three-state limit. (b) is a visualization of the B800 ring in LH-II \cite{cheng2006}, representing a specific application of the $N$-state spin-boson to modeling multi-chromophoric energy transfer in periodic systems.
    }
\end{figure}
\normalsize

\textit{Extensions.---}Having examined the single-mode case, the dissipative version is now defined. As an extension of Leggett {\it et al.} \cite{leggett}, consider a continuous $N$-well system with symmetric potential $V(\hat{X}, \hat{P})$ where the dynamics is restricted to the $N$-dimensional subspace of the well ground states. One then obtains $H_{sys}$ by introducing tunneling matrix elements $J_k$ between the wells [see Fig. \ref{f3}(a) for $N=3$]. With the dissipative environment approximated by a continuum of modes $\{\hat{q}_l,\hat{p}_l\}$, the $N$-state spin-boson Hamiltonian is simply Eq. (\ref{hd}) with $\{b, \omega,\ccc\} \rightarrow \{ b_l, \omega_l, \ccc_l\}$. The interaction term satisfies the criteria of \cite{setup_leggett_1983} and simplifies to the degenerate two-site spin-boson model at $N=2$. Other continuum normal modes can be added in a similar fashion \footnotemark[3].

Since the $N$-state model preserves rotational symmetry, the dissipative $H$ is an effective model for the single excitation manifold of a molecular ring or periodic chain interacting with a normal mode of a collective uncorrelated vibrational bath \cite{cheng2009}. This model specifically includes the geometrical structure of the system, an important property in excitation energy transfer \cite{Tei2011}. Couplings $J_k$ between all sites in the ring are included, allowing one to model systems with interactions other than nearest-neighbor. This version can model photoexcitation dynamics of molecular trimers \cite{Egger1994, Seibt2008, *pub5} and larger rings. A specific example is the 8-9 member B800 ring of photosynthetic LH-II \cite{cheng2006}, illustrated in Fig. \ref{f3}(b). Recently developed methods \cite{Kolli2011} for spin-boson dynamics can readily be applied to reveal similar insight into many-site systems as previous approaches \cite{leggett, Nazir2009, *brumer_pachon, *Tong2011} have revealed in the simplest two-site case.

As a final note, instead of extending the number of modes (or even reservoirs \cite{Velizhanin2008, *Nicolin2011}), the $N$-state Rabi model can be extended to many $N$-state systems. This approach would be similar to previous extensions of the two-state case \cite{taviscummings, *Tolkunov2007, *Koch2009a, *Agarwal2012}, but would include odd $N$, potentially revealing phase transitions and other interesting physics.

\textit{Summary.---}This work introduces an extension of the two-state Rabi model \cite{rabi, *rabi2} to describe dynamics of a more general $N$-state periodic system. The symmetry of the system is utilized in a group-theoretical approach, revealing insight into its energies and conserved quantities while also simplifying numerical analysis. A recently developed class of matrices \cite{gsm} provides an elegant method for obtaining the above results. Finally, the proposed infinite-mode extension generalizes the two-site spin-boson model \cite{leggett} to dissipative periodic $N$-site systems.

\begin{acknowledgments}
Discussions with M. H. Devoret, S. M. Girvin, F. Iachello, G. D. Scholes, and A. Nazir are gratefully acknowledged. I thank J. I. V\"ayrynen, K. A. Velizhanin, and D. Bokhan for help with preparation of this manuscript. This work is supported by an NSF Graduate Research Fellowship.
\end{acknowledgments}

\linespread{1}    
\bibliographystyle{apsrev4-1}
\bibliography{C:/Users/Anil_Smith/Documents/VVA_Documents/RESEARCH/Papers/library} 

\begin{thebibliography}{62}%
\makeatletter
\providecommand \@ifxundefined [1]{%
 \@ifx{#1\undefined}
}%
\providecommand \@ifnum [1]{%
 \ifnum #1\expandafter \@firstoftwo
 \else \expandafter \@secondoftwo
 \fi
}%
\providecommand \@ifx [1]{%
 \ifx #1\expandafter \@firstoftwo
 \else \expandafter \@secondoftwo
 \fi
}%
\providecommand \natexlab [1]{#1}%
\providecommand \enquote  [1]{``#1''}%
\providecommand \bibnamefont  [1]{#1}%
\providecommand \bibfnamefont [1]{#1}%
\providecommand \citenamefont [1]{#1}%
\providecommand \href@noop [0]{\@secondoftwo}%
\providecommand \href [0]{\begingroup \@sanitize@url \@href}%
\providecommand \@href[1]{\@@startlink{#1}\@@href}%
\providecommand \@@href[1]{\endgroup#1\@@endlink}%
\providecommand \@sanitize@url [0]{\catcode `\\12\catcode `\$12\catcode
  `\&12\catcode `\#12\catcode `\^12\catcode `\_12\catcode `\%12\relax}%
\providecommand \@@startlink[1]{}%
\providecommand \@@endlink[0]{}%
\providecommand \url  [0]{\begingroup\@sanitize@url \@url }%
\providecommand \@url [1]{\endgroup\@href {#1}{\urlprefix }}%
\providecommand \urlprefix  [0]{URL }%
\providecommand \Eprint [0]{\href }%
\providecommand \doibase [0]{http://dx.doi.org/}%
\providecommand \selectlanguage [0]{\@gobble}%
\providecommand \bibinfo  [0]{\@secondoftwo}%
\providecommand \bibfield  [0]{\@secondoftwo}%
\providecommand \translation [1]{[#1]}%
\providecommand \BibitemOpen [0]{}%
\providecommand \bibitemStop [0]{}%
\providecommand \bibitemNoStop [0]{.\EOS\space}%
\providecommand \EOS [0]{\spacefactor3000\relax}%
\providecommand \BibitemShut  [1]{\csname bibitem#1\endcsname}%
\let\auto@bib@innerbib\@empty
\bibitem [{\citenamefont {Rabi}(1936)}]{rabi}%
  \BibitemOpen
  \bibfield  {author} {\bibinfo {author} {\bibfnamefont {I.~I.}\ \bibnamefont
  {Rabi}},\ }\href {http://prola.aps.org/abstract/PR/v49/i4/p324\_1} {\bibfield
   {journal} {\bibinfo  {journal} {Phys. Rev.}\ }\textbf {\bibinfo {volume}
  {49}},\ \bibinfo {pages} {324} (\bibinfo {year} {1936})}\BibitemShut
  {NoStop}%
\bibitem [{\citenamefont {Rabi}(1937)}]{rabi2}%
  \BibitemOpen
  \bibfield  {author} {\bibinfo {author} {\bibfnamefont {I.~I.}\ \bibnamefont
  {Rabi}},\ }\href {\doibase 10.1103/PhysRev.51.652} {\bibfield  {journal}
  {\bibinfo  {journal} {Phys. Rev.}\ }\textbf {\bibinfo {volume} {51}},\
  \bibinfo {pages} {652} (\bibinfo {year} {1937})}\BibitemShut {NoStop}%
\bibitem [{\citenamefont {Fulton}\ and\ \citenamefont
  {Gouterman}(1961)}]{fg_original}%
  \BibitemOpen
  \bibfield  {author} {\bibinfo {author} {\bibfnamefont {R.~L.}\ \bibnamefont
  {Fulton}}\ and\ \bibinfo {author} {\bibfnamefont {M.}~\bibnamefont
  {Gouterman}},\ }\href {\doibase 10.1063/1.1701181} {\bibfield  {journal}
  {\bibinfo  {journal} {J. Chem. Phys.}\ }\textbf {\bibinfo {volume} {35}},\
  \bibinfo {pages} {1059} (\bibinfo {year} {1961})}\BibitemShut {NoStop}%
\bibitem [{\citenamefont {Wagner}(1984)}]{wagner_fg}%
  \BibitemOpen
  \bibfield  {author} {\bibinfo {author} {\bibfnamefont {M.}~\bibnamefont
  {Wagner}},\ }\href {\doibase doi:10.1088/0305-4470/17/11/026} {\bibfield
  {journal} {\bibinfo  {journal} {J. Phys. A: Math. Gen.}\ }\textbf {\bibinfo
  {volume} {17}},\ \bibinfo {pages} {2319} (\bibinfo {year}
  {1984})}\BibitemShut {NoStop}%
\bibitem [{\citenamefont {Jaynes}\ and\ \citenamefont {Cummings}(1963)}]{jc}%
  \BibitemOpen
  \bibfield  {author} {\bibinfo {author} {\bibfnamefont {E.~T.}\ \bibnamefont
  {Jaynes}}\ and\ \bibinfo {author} {\bibfnamefont {F.~W.}\ \bibnamefont
  {Cummings}},\ }\href {http://link.aps.org/doi/10.1103/PhysRevLett.78.3086}
  {\bibfield  {journal} {\bibinfo  {journal} {P. IEEE}\ }\textbf {\bibinfo
  {volume} {51}},\ \bibinfo {pages} {89} (\bibinfo {year} {1963})}\BibitemShut
  {NoStop}%
\bibitem [{\citenamefont {Paul}(1963)}]{Paul1963}%
  \BibitemOpen
  \bibfield  {author} {\bibinfo {author} {\bibfnamefont {H.}~\bibnamefont
  {Paul}},\ }\href {\doibase 10.1002/andp.19634660710} {\bibfield  {journal}
  {\bibinfo  {journal} {Ann. Phys. (Berlin)}\ }\textbf {\bibinfo {volume}
  {466}},\ \bibinfo {pages} {411} (\bibinfo {year} {1963})}\BibitemShut
  {NoStop}%
\bibitem [{\citenamefont {Shore}\ and\ \citenamefont
  {Knight}(1993)}]{Shore1993}%
  \BibitemOpen
  \bibfield  {author} {\bibinfo {author} {\bibfnamefont {B.~W.}\ \bibnamefont
  {Shore}}\ and\ \bibinfo {author} {\bibfnamefont {P.~L.}\ \bibnamefont
  {Knight}},\ }\href {\doibase 10.1080/09500349314551321} {\bibfield  {journal}
  {\bibinfo  {journal} {J. Mod. Optic.}\ }\textbf {\bibinfo {volume} {40}},\
  \bibinfo {pages} {1195} (\bibinfo {year} {1993})}\BibitemShut {NoStop}%
\bibitem [{\citenamefont {Shore}\ and\ \citenamefont
  {Sander}(1973)}]{spin_boson_shore_sander_FG}%
  \BibitemOpen
  \bibfield  {author} {\bibinfo {author} {\bibfnamefont {H.~B.}\ \bibnamefont
  {Shore}}\ and\ \bibinfo {author} {\bibfnamefont {L.~M.}\ \bibnamefont
  {Sander}},\ }\href {\doibase 10.1103/PhysRevB.7.4537} {\bibfield  {journal}
  {\bibinfo  {journal} {Phys. Rev. B}\ }\textbf {\bibinfo {volume} {7}},\
  \bibinfo {pages} {4537} (\bibinfo {year} {1973})}\BibitemShut {NoStop}%
\bibitem [{\citenamefont {Herfort}\ and\ \citenamefont
  {Wagner}(2001)}]{wagner}%
  \BibitemOpen
  \bibfield  {author} {\bibinfo {author} {\bibfnamefont {U.}~\bibnamefont
  {Herfort}}\ and\ \bibinfo {author} {\bibfnamefont {M.}~\bibnamefont
  {Wagner}},\ }\href {\doibase doi:10.1088/0953-8984/13/14/306} {\bibfield
  {journal} {\bibinfo  {journal} {J. Phys.: Condens. Matter}\ }\textbf
  {\bibinfo {volume} {13}},\ \bibinfo {pages} {3297} (\bibinfo {year}
  {2001})}\BibitemShut {NoStop}%
\bibitem [{\citenamefont {Viola}\ and\ \citenamefont
  {Lloyd}(1998)}]{Viola1998}%
  \BibitemOpen
  \bibfield  {author} {\bibinfo {author} {\bibfnamefont {L.}~\bibnamefont
  {Viola}}\ and\ \bibinfo {author} {\bibfnamefont {S.}~\bibnamefont {Lloyd}},\
  }\href {\doibase 10.1103/PhysRevA.58.2733} {\bibfield  {journal} {\bibinfo
  {journal} {Phys. Rev. A}\ }\textbf {\bibinfo {volume} {58}},\ \bibinfo
  {pages} {2733} (\bibinfo {year} {1998})}\BibitemShut {NoStop}%
\bibitem [{\citenamefont {Irish}\ \emph {et~al.}(2005)\citenamefont {Irish},
  \citenamefont {Gea-Banacloche}, \citenamefont {Martin},\ and\ \citenamefont
  {Schwab}}]{dimer_one_moder_small_j_irish}%
  \BibitemOpen
  \bibfield  {author} {\bibinfo {author} {\bibfnamefont {E.~K.}\ \bibnamefont
  {Irish}}, \bibinfo {author} {\bibfnamefont {J.}~\bibnamefont
  {Gea-Banacloche}}, \bibinfo {author} {\bibfnamefont {I.}~\bibnamefont
  {Martin}}, \ and\ \bibinfo {author} {\bibfnamefont {K.}~\bibnamefont
  {Schwab}},\ }\href {\doibase 10.1103/PhysRevB.72.195410} {\bibfield
  {journal} {\bibinfo  {journal} {Phys. Rev. B}\ }\textbf {\bibinfo {volume}
  {72}},\ \bibinfo {pages} {195410} (\bibinfo {year} {2005})}\BibitemShut
  {NoStop}%
\bibitem [{\citenamefont {Irish}(2007)}]{dimer_one_mode_irish_grwa_prl}%
  \BibitemOpen
  \bibfield  {author} {\bibinfo {author} {\bibfnamefont {E.~K.}\ \bibnamefont
  {Irish}},\ }\href {\doibase 10.1103/PhysRevLett.99.173601} {\bibfield
  {journal} {\bibinfo  {journal} {Phys. Rev. Lett.}\ }\textbf {\bibinfo
  {volume} {99}},\ \bibinfo {pages} {173601} (\bibinfo {year}
  {2007})}\BibitemShut {NoStop}%
\bibitem [{\citenamefont {Braak}(2011)}]{braak}%
  \BibitemOpen
  \bibfield  {author} {\bibinfo {author} {\bibfnamefont {D.}~\bibnamefont
  {Braak}},\ }\href {http://prl.aps.org/abstract/PRL/v107/i10/e100401}
  {\bibfield  {journal} {\bibinfo  {journal} {Phys. Rev. Lett.}\ }\textbf
  {\bibinfo {volume} {107}},\ \bibinfo {pages} {100401} (\bibinfo {year}
  {2011})}\BibitemShut {NoStop}%
\bibitem [{\citenamefont {Albert}\ \emph
  {et~al.}(2011{\natexlab{a}})\citenamefont {Albert}, \citenamefont {Scholes},\
  and\ \citenamefont {Brumer}}]{me}%
  \BibitemOpen
  \bibfield  {author} {\bibinfo {author} {\bibfnamefont {V.~V.}\ \bibnamefont
  {Albert}}, \bibinfo {author} {\bibfnamefont {G.~D.}\ \bibnamefont {Scholes}},
  \ and\ \bibinfo {author} {\bibfnamefont {P.}~\bibnamefont {Brumer}},\ }\href
  {http://pra.aps.org/abstract/PRA/v84/i4/e042110} {\bibfield  {journal}
  {\bibinfo  {journal} {Phys. Rev. A}\ }\textbf {\bibinfo {volume} {84}},\
  \bibinfo {pages} {042110} (\bibinfo {year} {2011}{\natexlab{a}})}\BibitemShut
  {NoStop}%
\bibitem [{\citenamefont {Scully}\ and\ \citenamefont
  {Zubairy}(1997)}]{scully}%
  \BibitemOpen
  \bibfield  {author} {\bibinfo {author} {\bibfnamefont {M.~O.}\ \bibnamefont
  {Scully}}\ and\ \bibinfo {author} {\bibfnamefont {M.~S.}\ \bibnamefont
  {Zubairy}},\ }\href@noop {} {\emph {\bibinfo {title} {{Quantum Optics}}}}\
  (\bibinfo  {publisher} {Cambridge University Press},\ \bibinfo {address}
  {Cambridge},\ \bibinfo {year} {1997})\BibitemShut {NoStop}%
\bibitem [{\citenamefont {Casanova}\ \emph {et~al.}(2010)\citenamefont
  {Casanova}, \citenamefont {Romero}, \citenamefont {Lizuain}, \citenamefont
  {Garc\'{\i}a-Ripoll},\ and\ \citenamefont
  {Solano}}]{spin_boson_deep_strong_coupling}%
  \BibitemOpen
  \bibfield  {author} {\bibinfo {author} {\bibfnamefont {J.}~\bibnamefont
  {Casanova}}, \bibinfo {author} {\bibfnamefont {G.}~\bibnamefont {Romero}},
  \bibinfo {author} {\bibfnamefont {I.}~\bibnamefont {Lizuain}}, \bibinfo
  {author} {\bibfnamefont {J.~J.}\ \bibnamefont {Garc\'{\i}a-Ripoll}}, \ and\
  \bibinfo {author} {\bibfnamefont {E.}~\bibnamefont {Solano}},\ }\href
  {\doibase 10.1103/PhysRevLett.105.263603} {\bibfield  {journal} {\bibinfo
  {journal} {Phys. Rev. Lett.}\ }\textbf {\bibinfo {volume} {105}},\ \bibinfo
  {pages} {263603} (\bibinfo {year} {2010})}\BibitemShut {NoStop}%
\bibitem [{\citenamefont {Larson}(2012)}]{Larson2011}%
  \BibitemOpen
  \bibfield  {author} {\bibinfo {author} {\bibfnamefont {J.}~\bibnamefont
  {Larson}},\ }\href {\doibase 10.1103/PhysRevLett.108.033601} {\bibfield
  {journal} {\bibinfo  {journal} {Phys. Rev. Lett.}\ }\textbf {\bibinfo
  {volume} {108}},\ \bibinfo {pages} {033601} (\bibinfo {year}
  {2012})}\BibitemShut {NoStop}%
\bibitem [{\citenamefont {You}\ and\ \citenamefont {Nori}(2011)}]{You2011}%
  \BibitemOpen
  \bibfield  {author} {\bibinfo {author} {\bibfnamefont {J.~Q.}\ \bibnamefont
  {You}}\ and\ \bibinfo {author} {\bibfnamefont {F.}~\bibnamefont {Nori}},\
  }\href {\doibase 10.1038/nature10122} {\bibfield  {journal} {\bibinfo
  {journal} {Nature}\ }\textbf {\bibinfo {volume} {474}},\ \bibinfo {pages}
  {589} (\bibinfo {year} {2011})}\BibitemShut {NoStop}%
\bibitem [{\citenamefont {Schuster}\ \emph {et~al.}(2010)\citenamefont
  {Schuster}, \citenamefont {Fragner}, \citenamefont {Dykman}, \citenamefont
  {Lyon},\ and\ \citenamefont {Schoelkopf}}]{Schuster2010}%
  \BibitemOpen
  \bibfield  {author} {\bibinfo {author} {\bibfnamefont {D.}~\bibnamefont
  {Schuster}}, \bibinfo {author} {\bibfnamefont {A.}~\bibnamefont {Fragner}},
  \bibinfo {author} {\bibfnamefont {M.}~\bibnamefont {Dykman}}, \bibinfo
  {author} {\bibfnamefont {S.}~\bibnamefont {Lyon}}, \ and\ \bibinfo {author}
  {\bibfnamefont {R.}~\bibnamefont {Schoelkopf}},\ }\href {\doibase
  10.1103/PhysRevLett.105.040503} {\bibfield  {journal} {\bibinfo  {journal}
  {Phys. Rev. Lett.}\ }\textbf {\bibinfo {volume} {105}},\ \bibinfo {pages}
  {040503} (\bibinfo {year} {2010})}\BibitemShut {NoStop}%
\bibitem [{\citenamefont {Crespi}\ \emph {et~al.}()\citenamefont {Crespi},
  \citenamefont {Longhi},\ and\ \citenamefont {Osellame}}]{Crespi2011}%
  \BibitemOpen
  \bibfield  {author} {\bibinfo {author} {\bibfnamefont {A.}~\bibnamefont
  {Crespi}}, \bibinfo {author} {\bibfnamefont {S.}~\bibnamefont {Longhi}}, \
  and\ \bibinfo {author} {\bibfnamefont {R.}~\bibnamefont {Osellame}},\ }\href
  {http://arxiv.org/abs/1111.6424} {\bibfield  {journal} {\bibinfo  {journal}
  {e-print}\ }}\Eprint {http://arxiv.org/abs/1111.6424} {arXiv:1111.6424}
  \BibitemShut {NoStop}%
\bibitem [{\citenamefont {Chang}\ \emph {et~al.}()\citenamefont {Chang},
  \citenamefont {Jiang}, \citenamefont {Gorshkov},\ and\ \citenamefont
  {Kimble}}]{Chang2012}%
  \BibitemOpen
  \bibfield  {author} {\bibinfo {author} {\bibfnamefont {D.~E.}\ \bibnamefont
  {Chang}}, \bibinfo {author} {\bibfnamefont {L.}~\bibnamefont {Jiang}},
  \bibinfo {author} {\bibfnamefont {A.~V.}\ \bibnamefont {Gorshkov}}, \ and\
  \bibinfo {author} {\bibfnamefont {H.~J.}\ \bibnamefont {Kimble}},\ }\href
  {http://arxiv.org/abs/1201.0643} {\bibfield  {journal} {\bibinfo  {journal}
  {e-print}\ }}\Eprint {http://arxiv.org/abs/1201.0643} {arXiv:1201.0643}
  \BibitemShut {NoStop}%
\bibitem [{\citenamefont {Longhi}(2011)}]{Longhi2011}%
  \BibitemOpen
  \bibfield  {author} {\bibinfo {author} {\bibfnamefont {S.}~\bibnamefont
  {Longhi}},\ }\href {http://www.ncbi.nlm.nih.gov/pubmed/21886226} {\bibfield
  {journal} {\bibinfo  {journal} {Opt. Lett.}\ }\textbf {\bibinfo {volume}
  {36}},\ \bibinfo {pages} {3407} (\bibinfo {year} {2011})}\BibitemShut
  {NoStop}%
\bibitem [{\citenamefont {Romero}\ \emph {et~al.}()\citenamefont {Romero},
  \citenamefont {Ballester}, \citenamefont {Wang}, \citenamefont {Scarani},\
  and\ \citenamefont {Solano}}]{Romero2012}%
  \BibitemOpen
  \bibfield  {author} {\bibinfo {author} {\bibfnamefont {G.}~\bibnamefont
  {Romero}}, \bibinfo {author} {\bibfnamefont {D.}~\bibnamefont {Ballester}},
  \bibinfo {author} {\bibfnamefont {Y.~M.}\ \bibnamefont {Wang}}, \bibinfo
  {author} {\bibfnamefont {V.}~\bibnamefont {Scarani}}, \ and\ \bibinfo
  {author} {\bibfnamefont {E.}~\bibnamefont {Solano}},\ }\href
  {http://arxiv.org/abs/1110.0223} {\bibfield  {journal} {\bibinfo  {journal}
  {e-print}\ }}\Eprint {http://arxiv.org/abs/1110.0223} {arXiv:1110.0223}
  \BibitemShut {NoStop}%
\bibitem [{\citenamefont {Scully}\ \emph {et~al.}(2006)\citenamefont {Scully},
  \citenamefont {Fry}, \citenamefont {Ooi},\ and\ \citenamefont
  {W\'{o}dkiewicz}}]{Scully2006}%
  \BibitemOpen
  \bibfield  {author} {\bibinfo {author} {\bibfnamefont {M.}~\bibnamefont
  {Scully}}, \bibinfo {author} {\bibfnamefont {E.}~\bibnamefont {Fry}},
  \bibinfo {author} {\bibfnamefont {C.}~\bibnamefont {Ooi}}, \ and\ \bibinfo
  {author} {\bibfnamefont {K.}~\bibnamefont {W\'{o}dkiewicz}},\ }\href
  {\doibase 10.1103/PhysRevLett.96.010501} {\bibfield  {journal} {\bibinfo
  {journal} {Phys. Rev. Lett.}\ }\textbf {\bibinfo {volume} {96}},\ \bibinfo
  {pages} {010501} (\bibinfo {year} {2006})}\BibitemShut {NoStop}%
\bibitem [{\citenamefont {Bianchetti}\ \emph {et~al.}(2010)\citenamefont
  {Bianchetti}, \citenamefont {Filipp}, \citenamefont {Baur}, \citenamefont
  {Fink}, \citenamefont {Lang}, \citenamefont {Steffen}, \citenamefont
  {Boissonneault}, \citenamefont {Blais},\ and\ \citenamefont
  {Wallraff}}]{Bianchetti2010}%
  \BibitemOpen
  \bibfield  {author} {\bibinfo {author} {\bibfnamefont {R.}~\bibnamefont
  {Bianchetti}}, \bibinfo {author} {\bibfnamefont {S.}~\bibnamefont {Filipp}},
  \bibinfo {author} {\bibfnamefont {M.}~\bibnamefont {Baur}}, \bibinfo {author}
  {\bibfnamefont {J.}~\bibnamefont {Fink}}, \bibinfo {author} {\bibfnamefont
  {C.}~\bibnamefont {Lang}}, \bibinfo {author} {\bibfnamefont {L.}~\bibnamefont
  {Steffen}}, \bibinfo {author} {\bibfnamefont {M.}~\bibnamefont
  {Boissonneault}}, \bibinfo {author} {\bibfnamefont {A.}~\bibnamefont
  {Blais}}, \ and\ \bibinfo {author} {\bibfnamefont {A.}~\bibnamefont
  {Wallraff}},\ }\href {http://prl.aps.org/abstract/PRL/v105/i22/e223601}
  {\bibfield  {journal} {\bibinfo  {journal} {Phys. Rev. Lett.}\ }\textbf
  {\bibinfo {volume} {105}},\ \bibinfo {pages} {223601} (\bibinfo {year}
  {2010})}\BibitemShut {NoStop}%
\bibitem [{\citenamefont {Srinivasan}\ \emph {et~al.}(2011)\citenamefont
  {Srinivasan}, \citenamefont {Hoffman}, \citenamefont {Gambetta},\ and\
  \citenamefont {Houck}}]{Srinivasan2011}%
  \BibitemOpen
  \bibfield  {author} {\bibinfo {author} {\bibfnamefont {S.}~\bibnamefont
  {Srinivasan}}, \bibinfo {author} {\bibfnamefont {A.}~\bibnamefont {Hoffman}},
  \bibinfo {author} {\bibfnamefont {J.}~\bibnamefont {Gambetta}}, \ and\
  \bibinfo {author} {\bibfnamefont {A.}~\bibnamefont {Houck}},\ }\href
  {\doibase 10.1103/PhysRevLett.106.083601} {\bibfield  {journal} {\bibinfo
  {journal} {Phys. Rev. Lett.}\ }\textbf {\bibinfo {volume} {106}},\ \bibinfo
  {pages} {083601} (\bibinfo {year} {2011})}\BibitemShut {NoStop}%
\bibitem [{\citenamefont {Shnyrkov}\ \emph {et~al.}()\citenamefont {Shnyrkov},
  \citenamefont {Soroka},\ and\ \citenamefont {Turutanov}}]{squitrid}%
  \BibitemOpen
  \bibfield  {author} {\bibinfo {author} {\bibfnamefont {V.~I.}\ \bibnamefont
  {Shnyrkov}}, \bibinfo {author} {\bibfnamefont {A.~A.}\ \bibnamefont
  {Soroka}}, \ and\ \bibinfo {author} {\bibfnamefont {O.~G.}\ \bibnamefont
  {Turutanov}},\ }\href {http://arxiv.org/abs/1111.6571} {\bibfield  {journal}
  {\bibinfo  {journal} {e-print}\ }}\Eprint {http://arxiv.org/abs/1111.6571}
  {arXiv:1111.6571} \BibitemShut {NoStop}%
\bibitem [{\citenamefont {Kolli}\ \emph {et~al.}(2011)\citenamefont {Kolli},
  \citenamefont {Nazir},\ and\ \citenamefont {Olaya-Castro}}]{Kolli2011}%
  \BibitemOpen
  \bibfield  {author} {\bibinfo {author} {\bibfnamefont {A.}~\bibnamefont
  {Kolli}}, \bibinfo {author} {\bibfnamefont {A.}~\bibnamefont {Nazir}}, \ and\
  \bibinfo {author} {\bibfnamefont {A.}~\bibnamefont {Olaya-Castro}},\ }\href
  {\doibase 10.1063/1.3652227} {\bibfield  {journal} {\bibinfo  {journal} {J.
  Chem. Phys.}\ }\textbf {\bibinfo {volume} {135}},\ \bibinfo {pages} {154112}
  (\bibinfo {year} {2011})}\BibitemShut {NoStop}%
\bibitem [{\citenamefont {Yoo}\ and\ \citenamefont {Eberly}(1985)}]{Yoo1985}%
  \BibitemOpen
  \bibfield  {author} {\bibinfo {author} {\bibfnamefont {H.-I.}\ \bibnamefont
  {Yoo}}\ and\ \bibinfo {author} {\bibfnamefont {J.~H.}\ \bibnamefont
  {Eberly}},\ }\href {\doibase 10.1016/0370-1573(85)90015-8} {\bibfield
  {journal} {\bibinfo  {journal} {Physics Reports}\ }\textbf {\bibinfo {volume}
  {118}},\ \bibinfo {pages} {239} (\bibinfo {year} {1985})}\BibitemShut
  {NoStop}%
\bibitem [{\citenamefont {Abdel-Hafez}\ \emph {et~al.}(1987)\citenamefont
  {Abdel-Hafez}, \citenamefont {Obada},\ and\ \citenamefont {Ahmad}}]{genjc3}%
  \BibitemOpen
  \bibfield  {author} {\bibinfo {author} {\bibfnamefont {A.~M.}\ \bibnamefont
  {Abdel-Hafez}}, \bibinfo {author} {\bibfnamefont {A.-S.~F.}\ \bibnamefont
  {Obada}}, \ and\ \bibinfo {author} {\bibfnamefont {M.~M.~A.}\ \bibnamefont
  {Ahmad}},\ }\href {\doibase 10.1103/PhysRevA.35.1634} {\bibfield  {journal}
  {\bibinfo  {journal} {Phys. Rev. A}\ }\textbf {\bibinfo {volume} {35}},\
  \bibinfo {pages} {1634} (\bibinfo {year} {1987})}\BibitemShut {NoStop}%
\bibitem [{\citenamefont {Hagelstein}\ and\ \citenamefont
  {Chaudhary}(2008)}]{Hagelstein2008}%
  \BibitemOpen
  \bibfield  {author} {\bibinfo {author} {\bibfnamefont {P.~L.}\ \bibnamefont
  {Hagelstein}}\ and\ \bibinfo {author} {\bibfnamefont {I.~U.}\ \bibnamefont
  {Chaudhary}},\ }\href {\doibase 10.1088/0953-4075/41/3/035601} {\bibfield
  {journal} {\bibinfo  {journal} {J. Phys. B: At. Mol. Opt. Phys.}\ }\textbf
  {\bibinfo {volume} {41}},\ \bibinfo {pages} {035601} (\bibinfo {year}
  {2008})}\BibitemShut {NoStop}%
\bibitem [{\citenamefont {Abdel-Wahab}(2011)}]{quadrumer_general}%
  \BibitemOpen
  \bibfield  {author} {\bibinfo {author} {\bibfnamefont {N.~H.}\ \bibnamefont
  {Abdel-Wahab}},\ }\href {\doibase 10.1140/epjp/i2011-11027-7} {\bibfield
  {journal} {\bibinfo  {journal} {Eur. Phys. J. Plus}\ }\textbf {\bibinfo
  {volume} {126}},\ \bibinfo {pages} {1} (\bibinfo {year} {2011})}\BibitemShut
  {NoStop}%
\bibitem [{\citenamefont {Wagner}\ and\ \citenamefont
  {Köngeter}(1989)}]{wagner_fg3}%
  \BibitemOpen
  \bibfield  {author} {\bibinfo {author} {\bibfnamefont {M.}~\bibnamefont
  {Wagner}}\ and\ \bibinfo {author} {\bibfnamefont {A.}~\bibnamefont
  {Köngeter}},\ }\href {\doibase 10.1063/1.456925} {\bibfield  {journal}
  {\bibinfo  {journal} {J. Chem. Phys.}\ }\textbf {\bibinfo {volume} {91}},\
  \bibinfo {pages} {3036} (\bibinfo {year} {1989})}\BibitemShut {NoStop}%
\bibitem [{\citenamefont {Eiermann}\ and\ \citenamefont
  {Wagner}(1996)}]{wagner_trimer}%
  \BibitemOpen
  \bibfield  {author} {\bibinfo {author} {\bibfnamefont {H.}~\bibnamefont
  {Eiermann}}\ and\ \bibinfo {author} {\bibfnamefont {M.}~\bibnamefont
  {Wagner}},\ }\href {\doibase 10.1063/1.471851} {\bibfield  {journal}
  {\bibinfo  {journal} {J. Chem. Phys.}\ }\textbf {\bibinfo {volume} {105}},\
  \bibinfo {pages} {6713} (\bibinfo {year} {1996})}\BibitemShut {NoStop}%
\bibitem [{\citenamefont {Rapp}\ and\ \citenamefont
  {Wagner}(1997)}]{wagner_gfg}%
  \BibitemOpen
  \bibfield  {author} {\bibinfo {author} {\bibfnamefont {M.}~\bibnamefont
  {Rapp}}\ and\ \bibinfo {author} {\bibfnamefont {M.}~\bibnamefont {Wagner}},\
  }\href {\doibase 10.1088/0305-4470/30/8/025} {\bibfield  {journal} {\bibinfo
  {journal} {J. Phys. A: Math. Gen.}\ }\textbf {\bibinfo {volume} {30}},\
  \bibinfo {pages} {2811} (\bibinfo {year} {1997})}\BibitemShut {NoStop}%
\bibitem [{\citenamefont {Leggett}\ \emph {et~al.}(1987)\citenamefont
  {Leggett}, \citenamefont {Chakravarty}, \citenamefont {Dorsey}, \citenamefont
  {Fisher}, \citenamefont {Garg},\ and\ \citenamefont {Zwerger}}]{leggett}%
  \BibitemOpen
  \bibfield  {author} {\bibinfo {author} {\bibfnamefont {A.~J.}\ \bibnamefont
  {Leggett}}, \bibinfo {author} {\bibfnamefont {S.}~\bibnamefont
  {Chakravarty}}, \bibinfo {author} {\bibfnamefont {A.~T.}\ \bibnamefont
  {Dorsey}}, \bibinfo {author} {\bibfnamefont {M.~P.~A.}\ \bibnamefont
  {Fisher}}, \bibinfo {author} {\bibfnamefont {A.}~\bibnamefont {Garg}}, \ and\
  \bibinfo {author} {\bibfnamefont {W.}~\bibnamefont {Zwerger}},\ }\href
  {http://rmp.aps.org/abstract/RMP/v59/i1/p1\_1} {\bibfield  {journal}
  {\bibinfo  {journal} {Rev. Mod. Phys.}\ }\textbf {\bibinfo {volume} {59}},\
  \bibinfo {pages} {1} (\bibinfo {year} {1987})}\BibitemShut {NoStop}%
\bibitem [{\citenamefont {Gilmore}\ and\ \citenamefont
  {McKenzie}(2005)}]{Gilmore2005}%
  \BibitemOpen
  \bibfield  {author} {\bibinfo {author} {\bibfnamefont {J.}~\bibnamefont
  {Gilmore}}\ and\ \bibinfo {author} {\bibfnamefont {R.~H.}\ \bibnamefont
  {McKenzie}},\ }\href {http://arxiv.org/abs/cond-mat/0401444} {\bibfield
  {journal} {\bibinfo  {journal} {J. Phys.: Condens. Matter}\ }\textbf
  {\bibinfo {volume} {17}},\ \bibinfo {pages} {1735} (\bibinfo {year}
  {2005})}\BibitemShut {NoStop}%
\bibitem [{\citenamefont {Gilmore}\ and\ \citenamefont
  {McKenzie}(2008)}]{Gilmore2008}%
  \BibitemOpen
  \bibfield  {author} {\bibinfo {author} {\bibfnamefont {J.}~\bibnamefont
  {Gilmore}}\ and\ \bibinfo {author} {\bibfnamefont {R.~H.}\ \bibnamefont
  {McKenzie}},\ }\href {http://dx.doi.org/10.1021/jp710243t} {\bibfield
  {journal} {\bibinfo  {journal} {J. Phys. Chem. A}\ }\textbf {\bibinfo
  {volume} {112}},\ \bibinfo {pages} {2162} (\bibinfo {year}
  {2008})}\BibitemShut {NoStop}%
\bibitem [{\citenamefont {Cheng}\ and\ \citenamefont
  {Fleming}(2009)}]{cheng2009}%
  \BibitemOpen
  \bibfield  {author} {\bibinfo {author} {\bibfnamefont {Y.-C.}\ \bibnamefont
  {Cheng}}\ and\ \bibinfo {author} {\bibfnamefont {G.~R.}\ \bibnamefont
  {Fleming}},\ }\href {\doibase 10.1146/annurev.physchem.040808.090259}
  {\bibfield  {journal} {\bibinfo  {journal} {Annu. Rev. Phys. Chem.}\ }\textbf
  {\bibinfo {volume} {60}},\ \bibinfo {pages} {241} (\bibinfo {year}
  {2009})}\BibitemShut {NoStop}%
\bibitem [{\citenamefont {Egger}\ and\ \citenamefont {Mak}(1994)}]{Egger1994}%
  \BibitemOpen
  \bibfield  {author} {\bibinfo {author} {\bibfnamefont {R.}~\bibnamefont
  {Egger}}\ and\ \bibinfo {author} {\bibfnamefont {C.~H.}\ \bibnamefont
  {Mak}},\ }\href {\doibase 10.1021/j100090a027} {\bibfield  {journal}
  {\bibinfo  {journal} {J. Phys. Chem.}\ }\textbf {\bibinfo {volume} {98}},\
  \bibinfo {pages} {9903} (\bibinfo {year} {1994})}\BibitemShut {NoStop}%
\bibitem [{\citenamefont {Jing}\ and\ \citenamefont {Yu}(2010)}]{Jing2010}%
  \BibitemOpen
  \bibfield  {author} {\bibinfo {author} {\bibfnamefont {J.}~\bibnamefont
  {Jing}}\ and\ \bibinfo {author} {\bibfnamefont {T.}~\bibnamefont {Yu}},\
  }\href {\doibase 10.1103/PhysRevLett.105.240403} {\bibfield  {journal}
  {\bibinfo  {journal} {Phys. Rev. Lett.}\ }\textbf {\bibinfo {volume} {105}},\
  \bibinfo {pages} {240403} (\bibinfo {year} {2010})}\BibitemShut {NoStop}%
\bibitem [{\citenamefont {Pittenger}\ and\ \citenamefont {Rubin}(2004)}]{gsm}%
  \BibitemOpen
  \bibfield  {author} {\bibinfo {author} {\bibfnamefont {A.}~\bibnamefont
  {Pittenger}}\ and\ \bibinfo {author} {\bibfnamefont {M.~H.}\ \bibnamefont
  {Rubin}},\ }\href {\doibase 10.1016/j.laa.2004.04.025} {\bibfield  {journal}
  {\bibinfo  {journal} {Linear Algebra Appl.}\ }\textbf {\bibinfo {volume}
  {390}},\ \bibinfo {pages} {255} (\bibinfo {year} {2004})}\BibitemShut
  {NoStop}%
\bibitem [{Note1()}]{Note1}%
  \BibitemOpen
  \bibinfo {note} {The main assumption of the model could in principle be
  manifested by aligning the respective dipole moments with the fields in such
  a way that the couplings contribute equally. Additionally, purely electric
  and magnetic dipole forms of $H_{int}$ are equivalent for the two-level case,
  as is the general rule [A. E. Siegman, {\protect \it Lasers} (University
  Science Books, Mill Valley, California, 1986) p. 1222]. This approach
  predicts that this is not the case for higher-level systems.}\BibitemShut
  {Stop}%
\bibitem [{Note2()}]{Note2}%
  \BibitemOpen
  \bibinfo {note} {See Supplemental Material at [URL] for details on the
  generalized spin matrices, the Fulton-Gouterman \cite {fg_original,
  *wagner_fg} transformation that partially diagonalizes $H$, and the
  S-GRWA.}\BibitemShut {Stop}%
\bibitem [{Note3()}]{Note3}%
  \BibitemOpen
  \bibinfo {note} {For the three-level two-mode case, another mode $b_2$ could
  be added to $H_3$ via $b_2 S_{2,0} + b_2^\dagger S^\dagger _{2,0}$. One can
  compare to a cascade ($\Xi $) three-level configuration, recently expressed
  in the position representation in X. Ren, H. Cong, X. Wang, and J. Xia, Sci.
  China Phys. Mech. Astron. {\protect \bf 54}, 1625 (2011). For arbitrary $N$,
  a mode $b_n$ can couple via $S_{n,0}$ for $n<N$, providing a non-RWA model of
  interaction between an $N$-state atom and $N-1$ modes \cite {genjc3} which
  preserves cyclic symmetry.}\BibitemShut {Stop}%
\bibitem [{\citenamefont {Boissonneault}\ \emph {et~al.}(2008)\citenamefont
  {Boissonneault}, \citenamefont {Gambetta},\ and\ \citenamefont
  {Blais}}]{Boissonneault2008}%
  \BibitemOpen
  \bibfield  {author} {\bibinfo {author} {\bibfnamefont {M.}~\bibnamefont
  {Boissonneault}}, \bibinfo {author} {\bibfnamefont {J.~M.}\ \bibnamefont
  {Gambetta}}, \ and\ \bibinfo {author} {\bibfnamefont {A.}~\bibnamefont
  {Blais}},\ }\href {\doibase 10.1103/PhysRevA.77.060305} {\bibfield  {journal}
  {\bibinfo  {journal} {Phys. Rev. A}\ }\textbf {\bibinfo {volume} {77}},\
  \bibinfo {pages} {060305(R)} (\bibinfo {year} {2008})}\BibitemShut {NoStop}%
\bibitem [{\citenamefont {Zueco}\ \emph {et~al.}(2009)\citenamefont {Zueco},
  \citenamefont {Reuther}, \citenamefont {Kohler},\ and\ \citenamefont
  {H\"{a}nggi}}]{Zueco2009}%
  \BibitemOpen
  \bibfield  {author} {\bibinfo {author} {\bibfnamefont {D.}~\bibnamefont
  {Zueco}}, \bibinfo {author} {\bibfnamefont {G.}~\bibnamefont {Reuther}},
  \bibinfo {author} {\bibfnamefont {S.}~\bibnamefont {Kohler}}, \ and\ \bibinfo
  {author} {\bibfnamefont {P.}~\bibnamefont {H\"{a}nggi}},\ }\href {\doibase
  10.1103/PhysRevA.80.033846} {\bibfield  {journal} {\bibinfo  {journal} {Phys.
  Rev. A}\ }\textbf {\bibinfo {volume} {80}},\ \bibinfo {pages} {033846}
  (\bibinfo {year} {2009})}\BibitemShut {NoStop}%
\bibitem [{\citenamefont {Fleischhauer}\ \emph {et~al.}(2005)\citenamefont
  {Fleischhauer}, \citenamefont {Imamoglu},\ and\ \citenamefont
  {Marangos}}]{Fleischhauer2005}%
  \BibitemOpen
  \bibfield  {author} {\bibinfo {author} {\bibfnamefont {M.}~\bibnamefont
  {Fleischhauer}}, \bibinfo {author} {\bibfnamefont {A.}~\bibnamefont
  {Imamoglu}}, \ and\ \bibinfo {author} {\bibfnamefont {J.}~\bibnamefont
  {Marangos}},\ }\href {\doibase 10.1103/RevModPhys.77.633} {\bibfield
  {journal} {\bibinfo  {journal} {Rev. Mod. Phys.}\ }\textbf {\bibinfo {volume}
  {77}},\ \bibinfo {pages} {633} (\bibinfo {year} {2005})}\BibitemShut
  {NoStop}%
\bibitem [{\citenamefont {Cheng}\ and\ \citenamefont
  {Silbey}(2006)}]{cheng2006}%
  \BibitemOpen
  \bibfield  {author} {\bibinfo {author} {\bibfnamefont {Y.-C.}\ \bibnamefont
  {Cheng}}\ and\ \bibinfo {author} {\bibfnamefont {R.}~\bibnamefont {Silbey}},\
  }\href {http://prl.aps.org/abstract/PRL/v96/i2/e028103} {\bibfield  {journal}
  {\bibinfo  {journal} {Phys. Rev. Lett.}\ }\textbf {\bibinfo {volume} {96}},\
  \bibinfo {pages} {028103} (\bibinfo {year} {2006})}\BibitemShut {NoStop}%
\bibitem [{\citenamefont {Caldeira}\ and\ \citenamefont
  {Leggett}(1983)}]{setup_leggett_1983}%
  \BibitemOpen
  \bibfield  {author} {\bibinfo {author} {\bibfnamefont {A.}~\bibnamefont
  {Caldeira}}\ and\ \bibinfo {author} {\bibfnamefont {A.~J.}\ \bibnamefont
  {Leggett}},\ }\href {\doibase 10.1016/0003-4916(83)90202-6} {\bibfield
  {journal} {\bibinfo  {journal} {Ann. Phys.}\ }\textbf {\bibinfo {volume}
  {149}},\ \bibinfo {pages} {374, Appendix C} (\bibinfo {year}
  {1983})}\BibitemShut {NoStop}%
\bibitem [{\citenamefont {Tei}\ \emph {et~al.}(2011)\citenamefont {Tei},
  \citenamefont {Nakatani},\ and\ \citenamefont {Ishihara}}]{Tei2011}%
  \BibitemOpen
  \bibfield  {author} {\bibinfo {author} {\bibfnamefont {G.}~\bibnamefont
  {Tei}}, \bibinfo {author} {\bibfnamefont {M.}~\bibnamefont {Nakatani}}, \
  and\ \bibinfo {author} {\bibfnamefont {H.}~\bibnamefont {Ishihara}},\ }\href
  {\doibase 10.1002/pssb.201000655} {\bibfield  {journal} {\bibinfo  {journal}
  {Phys. Status Solidi B}\ }\textbf {\bibinfo {volume} {248}},\ \bibinfo
  {pages} {399} (\bibinfo {year} {2011})}\BibitemShut {NoStop}%
\bibitem [{\citenamefont {Seibt}\ and\ \citenamefont
  {Engel}(2008)}]{Seibt2008}%
  \BibitemOpen
  \bibfield  {author} {\bibinfo {author} {\bibfnamefont {J.}~\bibnamefont
  {Seibt}}\ and\ \bibinfo {author} {\bibfnamefont {V.}~\bibnamefont {Engel}},\
  }\href {\doibase 10.1016/j.chemphys.2007.10.014} {\bibfield  {journal}
  {\bibinfo  {journal} {J. Chem. Phys.}\ }\textbf {\bibinfo {volume} {347}},\
  \bibinfo {pages} {120} (\bibinfo {year} {2008})}\BibitemShut {NoStop}%
\bibitem [{\citenamefont {Albert}\ \emph
  {et~al.}(2011{\natexlab{b}})\citenamefont {Albert}, \citenamefont {Badaeva},
  \citenamefont {Kilina}, \citenamefont {Sykora},\ and\ \citenamefont
  {Tretiak}}]{pub5}%
  \BibitemOpen
  \bibfield  {author} {\bibinfo {author} {\bibfnamefont {V.~V.}\ \bibnamefont
  {Albert}}, \bibinfo {author} {\bibfnamefont {E.}~\bibnamefont {Badaeva}},
  \bibinfo {author} {\bibfnamefont {S.}~\bibnamefont {Kilina}}, \bibinfo
  {author} {\bibfnamefont {M.}~\bibnamefont {Sykora}}, \ and\ \bibinfo {author}
  {\bibfnamefont {S.}~\bibnamefont {Tretiak}},\ }\href {\doibase
  10.1016/j.jlumin.2011.04.009} {\bibfield  {journal} {\bibinfo  {journal} {J.
  Lumin.}\ }\textbf {\bibinfo {volume} {131}},\ \bibinfo {pages} {1739}
  (\bibinfo {year} {2011}{\natexlab{b}})}\BibitemShut {NoStop}%
\bibitem [{\citenamefont {Nazir}(2009)}]{Nazir2009}%
  \BibitemOpen
  \bibfield  {author} {\bibinfo {author} {\bibfnamefont {A.}~\bibnamefont
  {Nazir}},\ }\href {\doibase 10.1103/PhysRevLett.103.146404} {\bibfield
  {journal} {\bibinfo  {journal} {Phys. Rev. Lett.}\ }\textbf {\bibinfo
  {volume} {103}},\ \bibinfo {pages} {146404} (\bibinfo {year}
  {2009})}\BibitemShut {NoStop}%
\bibitem [{\citenamefont {Pach\'{o}n}\ and\ \citenamefont
  {Brumer}(2011)}]{brumer_pachon}%
  \BibitemOpen
  \bibfield  {author} {\bibinfo {author} {\bibfnamefont {L.~A.}\ \bibnamefont
  {Pach\'{o}n}}\ and\ \bibinfo {author} {\bibfnamefont {P.}~\bibnamefont
  {Brumer}},\ }\href {\doibase 10.1021/jz201189p} {\bibfield  {journal}
  {\bibinfo  {journal} {J. Phys. Chem. Lett.}\ }\textbf {\bibinfo {volume}
  {2}},\ \bibinfo {pages} {2728} (\bibinfo {year} {2011})}\BibitemShut
  {NoStop}%
\bibitem [{\citenamefont {Tong}\ \emph {et~al.}(2011)\citenamefont {Tong},
  \citenamefont {An}, \citenamefont {Luo},\ and\ \citenamefont
  {Oh}}]{Tong2011}%
  \BibitemOpen
  \bibfield  {author} {\bibinfo {author} {\bibfnamefont {Q.-J.}\ \bibnamefont
  {Tong}}, \bibinfo {author} {\bibfnamefont {J.-H.}\ \bibnamefont {An}},
  \bibinfo {author} {\bibfnamefont {H.-G.}\ \bibnamefont {Luo}}, \ and\
  \bibinfo {author} {\bibfnamefont {C.}~\bibnamefont {Oh}},\ }\href {\doibase
  10.1103/PhysRevB.84.174301} {\bibfield  {journal} {\bibinfo  {journal} {Phys.
  Rev. B}\ }\textbf {\bibinfo {volume} {84}},\ \bibinfo {pages} {174301}
  (\bibinfo {year} {2011})}\BibitemShut {NoStop}%
\bibitem [{\citenamefont {Velizhanin}\ \emph {et~al.}(2008)\citenamefont
  {Velizhanin}, \citenamefont {Wang},\ and\ \citenamefont
  {Thoss}}]{Velizhanin2008}%
  \BibitemOpen
  \bibfield  {author} {\bibinfo {author} {\bibfnamefont {K.~A.}\ \bibnamefont
  {Velizhanin}}, \bibinfo {author} {\bibfnamefont {H.}~\bibnamefont {Wang}}, \
  and\ \bibinfo {author} {\bibfnamefont {M.}~\bibnamefont {Thoss}},\ }\href
  {\doibase 10.1016/j.cplett.2008.05.065} {\bibfield  {journal} {\bibinfo
  {journal} {Chem. Phys. Lett.}\ }\textbf {\bibinfo {volume} {460}},\ \bibinfo
  {pages} {325} (\bibinfo {year} {2008})}\BibitemShut {NoStop}%
\bibitem [{\citenamefont {Nicolin}\ and\ \citenamefont
  {Segal}(2011)}]{Nicolin2011}%
  \BibitemOpen
  \bibfield  {author} {\bibinfo {author} {\bibfnamefont {L.}~\bibnamefont
  {Nicolin}}\ and\ \bibinfo {author} {\bibfnamefont {D.}~\bibnamefont
  {Segal}},\ }\href {\doibase 10.1103/PhysRevB.84.161414} {\bibfield  {journal}
  {\bibinfo  {journal} {Phys. Rev. B}\ }\textbf {\bibinfo {volume} {84}},\
  \bibinfo {pages} {161414(R)} (\bibinfo {year} {2011})}\BibitemShut {NoStop}%
\bibitem [{\citenamefont {Tavis}\ and\ \citenamefont
  {Cummings}(1968)}]{taviscummings}%
  \BibitemOpen
  \bibfield  {author} {\bibinfo {author} {\bibfnamefont {M.}~\bibnamefont
  {Tavis}}\ and\ \bibinfo {author} {\bibfnamefont {F.}~\bibnamefont
  {Cummings}},\ }\href {\doibase 10.1103/PhysRev.170.379} {\bibfield  {journal}
  {\bibinfo  {journal} {Phys. Rev.}\ }\textbf {\bibinfo {volume} {170}},\
  \bibinfo {pages} {379} (\bibinfo {year} {1968})}\BibitemShut {NoStop}%
\bibitem [{\citenamefont {Tolkunov}\ and\ \citenamefont
  {Solenov}(2007)}]{Tolkunov2007}%
  \BibitemOpen
  \bibfield  {author} {\bibinfo {author} {\bibfnamefont {D.}~\bibnamefont
  {Tolkunov}}\ and\ \bibinfo {author} {\bibfnamefont {D.}~\bibnamefont
  {Solenov}},\ }\href {\doibase 10.1103/PhysRevB.75.024402} {\bibfield
  {journal} {\bibinfo  {journal} {Phys. Rev. B}\ }\textbf {\bibinfo {volume}
  {75}},\ \bibinfo {pages} {024402} (\bibinfo {year} {2007})}\BibitemShut
  {NoStop}%
\bibitem [{\citenamefont {Koch}\ and\ \citenamefont {{Le
  Hur}}(2009)}]{Koch2009a}%
  \BibitemOpen
  \bibfield  {author} {\bibinfo {author} {\bibfnamefont {J.}~\bibnamefont
  {Koch}}\ and\ \bibinfo {author} {\bibfnamefont {K.}~\bibnamefont {{Le
  Hur}}},\ }\href {\doibase 10.1103/PhysRevA.80.023811} {\bibfield  {journal}
  {\bibinfo  {journal} {Phys. Rev. A}\ }\textbf {\bibinfo {volume} {80}},\
  \bibinfo {pages} {023811} (\bibinfo {year} {2009})}\BibitemShut {NoStop}%
\bibitem [{\citenamefont {Agarwal}\ \emph {et~al.}()\citenamefont {Agarwal},
  \citenamefont {Rafsanjani},\ and\ \citenamefont {Eberly}}]{Agarwal2012}%
  \BibitemOpen
  \bibfield  {author} {\bibinfo {author} {\bibfnamefont {S.}~\bibnamefont
  {Agarwal}}, \bibinfo {author} {\bibfnamefont {S.~M.~H.}\ \bibnamefont
  {Rafsanjani}}, \ and\ \bibinfo {author} {\bibfnamefont {J.~H.}\ \bibnamefont
  {Eberly}},\ }\href {http://arxiv.org/abs/1201.2928} {\bibfield  {journal}
  {\bibinfo  {journal} {e-print}\ }}\Eprint {http://arxiv.org/abs/1201.2928}
  {arXiv:1201.2928} \BibitemShut {NoStop}%
\end{thebibliography}%


\begin{thebibliography}{3}%
\makeatletter
\providecommand \@ifxundefined [1]{%
 \@ifx{#1\undefined}
}%
\providecommand \@ifnum [1]{%
 \ifnum #1\expandafter \@firstoftwo
 \else \expandafter \@secondoftwo
 \fi
}%
\providecommand \@ifx [1]{%
 \ifx #1\expandafter \@firstoftwo
 \else \expandafter \@secondoftwo
 \fi
}%
\providecommand \natexlab [1]{#1}%
\providecommand \enquote  [1]{``#1''}%
\providecommand \bibnamefont  [1]{#1}%
\providecommand \bibfnamefont [1]{#1}%
\providecommand \citenamefont [1]{#1}%
\providecommand \href@noop [0]{\@secondoftwo}%
\providecommand \href [0]{\begingroup \@sanitize@url \@href}%
\providecommand \@href[1]{\@@startlink{#1}\@@href}%
\providecommand \@@href[1]{\endgroup#1\@@endlink}%
\providecommand \@sanitize@url [0]{\catcode `\\12\catcode `\$12\catcode
  `\&12\catcode `\#12\catcode `\^12\catcode `\_12\catcode `\%12\relax}%
\providecommand \@@startlink[1]{}%
\providecommand \@@endlink[0]{}%
\providecommand \url  [0]{\begingroup\@sanitize@url \@url }%
\providecommand \@url [1]{\endgroup\@href {#1}{\urlprefix }}%
\providecommand \urlprefix  [0]{URL }%
\providecommand \Eprint [0]{\href }%
\providecommand \doibase [0]{http://dx.doi.org/}%
\providecommand \selectlanguage [0]{\@gobble}%
\providecommand \bibinfo  [0]{\@secondoftwo}%
\providecommand \bibfield  [0]{\@secondoftwo}%
\providecommand \translation [1]{[#1]}%
\providecommand \BibitemOpen [0]{}%
\providecommand \bibitemStop [0]{}%
\providecommand \bibitemNoStop [0]{.\EOS\space}%
\providecommand \EOS [0]{\spacefactor3000\relax}%
\providecommand \BibitemShut  [1]{\csname bibitem#1\endcsname}%
\let\auto@bib@innerbib\@empty
\bibitem [{\citenamefont {Pittenger}\ and\ \citenamefont {Rubin}(2004)}]{gsm}%
  \BibitemOpen
  \bibfield  {author} {\bibinfo {author} {\bibfnamefont {A.}~\bibnamefont
  {Pittenger}}\ and\ \bibinfo {author} {\bibfnamefont {M.~H.}\ \bibnamefont
  {Rubin}},\ }\href {\doibase 10.1016/j.laa.2004.04.025} {\bibfield  {journal}
  {\bibinfo  {journal} {Linear Algebra Appl.}\ }\textbf {\bibinfo {volume}
  {390}},\ \bibinfo {pages} {255} (\bibinfo {year} {2004})}\BibitemShut
  {NoStop}%
\bibitem [{\citenamefont {Wagner}(1984)}]{wagner_fg}%
  \BibitemOpen
  \bibfield  {author} {\bibinfo {author} {\bibfnamefont {M.}~\bibnamefont
  {Wagner}},\ }\href {\doibase doi:10.1088/0305-4470/17/11/026} {\bibfield
  {journal} {\bibinfo  {journal} {J. Phys. A: Math. Gen.}\ }\textbf {\bibinfo
  {volume} {17}},\ \bibinfo {pages} {2319} (\bibinfo {year}
  {1984})}\BibitemShut {NoStop}%
\bibitem [{\citenamefont {Albert}\ \emph {et~al.}(2011)\citenamefont {Albert},
  \citenamefont {Scholes},\ and\ \citenamefont {Brumer}}]{me}%
  \BibitemOpen
  \bibfield  {author} {\bibinfo {author} {\bibfnamefont {V.~V.}\ \bibnamefont
  {Albert}}, \bibinfo {author} {\bibfnamefont {G.~D.}\ \bibnamefont {Scholes}},
  \ and\ \bibinfo {author} {\bibfnamefont {P.}~\bibnamefont {Brumer}},\ }\href
  {http://pra.aps.org/abstract/PRA/v84/i4/e042110} {\bibfield  {journal}
  {\bibinfo  {journal} {Phys. Rev. A}\ }\textbf {\bibinfo {volume} {84}},\
  \bibinfo {pages} {042110} (\bibinfo {year} {2011})}\BibitemShut {NoStop}%
\end{thebibliography}%

\end{document}


\title{Supplemental Material: A quantum Rabi model for $N$-state atoms}
\author{Victor V. Albert}
\email[]{victor.albert@yale.edu}
\affiliation{Department of Physics, Yale University, P.O. Box 208120, New Haven, CT 06520-8120, USA}
\begin{abstract}
\end{abstract}
\maketitle

\subsection*{Generalized spin matrices}

Suppressing dependence on $N$, generalized spin matrices \cite{gsm} for $0\leq j,k < N$ are once again defined (via modulo $N$) as
\begin{equation}\label{spin}
S_{j,k}=\sum_{n=0}^{N-1}e^{i\frac{2\pi}{N}nj}|n\ket\bra n+k|=\left(S_{1,0}\right)^{j}\left(S_{0,1}\right)^{k}.\nonumber
\end{equation}
The matrices $S_{1,0}$ and $S_{0,1}$ are the generators of this unitary set and $S_{0,0}$ is the identity. These matrices are ideal for describing periodic $N$-level systems where the site interactions are circulant. While $S^\dagger_{j,0}=S_{-j,0}$ and $S^\dagger_{0,k}=S_{0,-k}$, for arbitrary matrices $S^\dagger_{j,k} = e^{i\frac{2\pi}{N}jk}S_{-j,-k}$. For $N=2$, $S_{0,1}=\sigma_x$, $S_{1,0}=\sigma_z$, and $S_{1,1}=i\sigma_y$. For $N=3$, the relevant matrices are
\ba\nonumber
S_{0,0}&=&\left(\begin{array}{ccc}
1 & 0 & 0\\
0 & 1 & 0\\
0 & 0 & 1
\end{array}\right),\,\,\,\,\,\,\,\,\,\,\,\,\,\,\,\,\,\,\,\,\, S_{0,1}=\left(\begin{array}{ccc}
0 & 1 & 0\\
0 & 0 & 1\\
1 & 0 & 0
\end{array}\right),\,\,\,\,\,\,\,\,\,\,\,\,\,\,\,\,\,\,\,\,\, S_{0,2}=\left(\begin{array}{ccc}
0 & 0 & 1\\
1 & 0 & 0\\
0 & 1 & 0
\end{array}\right),\\\nonumber
S_{1,0}&=&\left(\begin{array}{ccc}
1 & 0 & 0\\
0 & e^{i\frac{2\pi}{3}} & 0\\
0 & 0 & e^{-i\frac{2\pi}{3}}
\end{array}\right),\,\, S_{1,1}=\left(\begin{array}{ccc}
0 & 1 & 0\\
0 & 0 & e^{i\frac{2\pi}{3}}\\
e^{-i\frac{2\pi}{3}} & 0 & 0
\end{array}\right),\,\, S_{1,2}=\left(\begin{array}{ccc}
0 & 0 & 1\\
e^{i\frac{2\pi}{3}} & 0 & 0\\
0 & e^{-i\frac{2\pi}{3}} & 0
\end{array}\right),\\S_{2,0}&=&\left(\begin{array}{ccc}
1 & 0 & 0\\
0 & e^{-i\frac{2\pi}{3}} & 0\\
0 & 0 & e^{i\frac{2\pi}{3}}
\end{array}\right),\,\, S_{2,1}=\left(\begin{array}{ccc}
0 & 1 & 0\\
0 & 0 & e^{-i\frac{2\pi}{3}}\\
e^{i\frac{2\pi}{3}} & 0 & 0
\end{array}\right),\,\, S_{2,2}=\left(\begin{array}{ccc}
0 & 0 & 1\\
e^{-i\frac{2\pi}{3}} & 0 & 0\\
0 & e^{i\frac{2\pi}{3}} & 0
\end{array}\right)\nonumber
.
\ea
Letting $a,b,j,k$ be integers, one obtains a phase when two matrices are permuted: $S_{a,b}S_{j,k}=e^{i\frac{2\pi}{N}\left(bj-ka\right)}S_{j,k}S_{a,b}$. An additional identity useful in manipulating these matrices and proving the steps below is
\be
\frac{1}{N}\sum_{n=0}^{N-1}e^{i\frac{2\pi}{N}n\left(l-m\right)}=\delta_{l,m}.\nonumber
\ee

\subsection{Partial diagonalization of the $N$-state Rabi model}

Since the $N$-state Rabi model contains an $N$-fold rotational symmetry, one can partially diagonalize $H$ in the site subspace using the prescription of \cite{wagner_fg}. The generalized spin matrices significantly simplify this prescription and the general transformation is a composition of spin and boson operators:
\begin{equation}\label{ud}
U=\frac{1}{\sqrt{N}}\sum_{k=0}^{N-1}e^{i\frac{2\pi}{N}k^{2}}\mathcal{R}_{k}S_{k,k}.\nonumber
\end{equation}
This unitary $U$ commutes with the coupling term $H_{int}$ since the phase contribution from permuting the spin matrices exactly cancels the phase from permuting the boson operators. The transformation partially diagonalizes the system matrix $H_{sys}$ since $U^{\dagger}S_{0,k}U=\mathcal{R}_{k}^{\dagger}S_{k,0}^{\dagger}$. The transformed $H$ is diagonal in the spin space with $\bra n^{\prime}|\widetilde{H}|n\ket=\delta_{n^{\prime},n}\widetilde{H}_{N,n}$ and boson chains
\be\label{hdt}
\nonumber
\widetilde{H}_{N,n}=\bra n|U^\dagger HU|n\ket=\omega b^{\dagger}b+\lambda(be^{i\frac{2\pi}{N}n}+b^{\dagger}e^{-i\frac{2\pi}{N}n})+\left(-1\right)^{n}J\mathcal{R}+\sum_{k=1}^{\kappa}J_{k}(\mathcal{R}_{k}e^{i\frac{2\pi}{N}nk}+\mathcal{R}_{k}^{\dagger}e^{-i\frac{2\pi}{N}nk}).
\ee
The transformed form of the commuting operator $\mathbf{N}$ is simply the diagonal form of $H_{sys}$, namely,
\be\nonumber
\widetilde{\mathbf{N}}=\widehat{H}_{sys}=JS_{\frac{N}{2},0}+\sum_{k=1}^{\kappa}J_{k}(S_{k,0}+S_{k,0}^{\dagger}).
\ee
For the $n$th diagonal of $\widetilde{\mathbf{N}}$, the coefficients of the site couplings $J_k$ are conserved quantum numbers that describe the corresponding chain $\widetilde{H}_{N,n}$. As an example, $\widetilde{\mathbf{N}}=J\sigma_z$ when $N=2$, revealing chains of parity $\pm1$. Likewise, $\widetilde{\mathbf{N}} = \text{diag}\left\{ 2K,-K,-K\right\}$ for $N=3$, revealing a conserved quantum number $\delta=-2,1$. For $N=4$, $\widetilde{\mathbf{N}} = \text{diag}\left\{ J+2K,-J,J-2K,-J\right\}$ and there are two quantum numbers: the parity $p=\pm$ and the cascade number $\delta=0,\pm2$. The former determines whether the state is in the degenerate $-J$ chain while the latter resolves the remaining two chains $J\pm2K$. Note that there are no quantum numbers which can be used to distinguish the negative parity chains for the four-state case.

\subsection{Symmetric generalized rotating wave approximation (S-GRWA)}

The 1-by-1 S-GRWA \cite{me} is derived by applying the generalized displacement operator
\begin{equation}\nonumber
\mathcal{D}(S_{1,0}\lambda/\omega)=\exp[(bS_{1,0}-b^{\dagger}S_{1,0}^{\dagger})\lambda/\omega]
\end{equation}
to $\widetilde{H}$ and writing out each transformed chain of $\mathcal{D}^\dagger \widetilde{H} \mathcal{D}$ in the boson Fock space. This removes the linear bosonic coupling terms and (keeping only diagonal elements) obtains energies (for $m=0,1,2,\ldots$)
\begin{widetext}
\begin{equation}\label{srwa}\nonumber
E_{N,n,m}^{\text{S-GRWA}}=\omega m-\frac{\lambda^{2}}{\omega}+Je^{-2\lambda^{2}/\omega^{2}}L_{m}(4\lambda^{2}/\omega^{2})(-1)^{n+m} +\sum_{k=1}^{\kappa}2J_{k}e^{-\half|\alpha_{k}\lambda/\omega|^{2}}L_{m}\left(\left|\alpha_{k}\lambda/\omega\right|^{2}\right)\cos\left(\frac{2\pi}{N}k\left(n+m\right)+\theta_{k}\right)
\nonumber
\end{equation}
\end{widetext}
where the third term is only for even $N$, $L_m(x)$ is a Laguerre polynomial, and
\begin{eqnarray}
\theta_{k}&=&\left(\lambda/\omega\right)^{2}\sin\left(2\pi k/N\right)\nonumber\\
\alpha_{k}&=&1-e^{i2\pi k/N}\nonumber.
\end{eqnarray}
For $N=2$, the 1-by-1 S-GRWA simplifies to well-known results (see \cite{me} and refs. therein). Interpreting $H$ as a tunneling $N$-site system in the position representation, the energies consist of contributions from standing waves along the ring/chain of sites weighted by site coupling $J_k$, displaced by $\theta_k$, and decaying as $\mathcal{O}(\lambda^2)$. The corresponding eigenfunctions consist of the position basis rotated along the ring by $U^\dagger$ and displaced vibrationally by $\mathcal{D}^\dagger$. While the opposite-site (parity) interaction $J$ is the only interaction in the two-state case, it naturally becomes less relevant as $2N$ becomes large. The two limits of the system are $\lambda \gg \{J_{k}\}$, which simplifies $H$ into a collection of symmetrically displaced harmonic oscillators, and $\lambda \ll \{J_{k}\}$, which results in an uncoupled $N$-state system and an oscillator.

\linespread{1}    
\bibliographystyle{apsrev4-1}
\bibliography{V:/Files/VVA_Documents/RESEARCH/Papers/library} 